\documentstyle[12pt]{article}
\def\strut{\rule[-.5cm]{0cm}{1cm}}

\def\dspace{\baselineskip = .30in}

\begin{document}

\title{Natural Gauge Hierarchy in SO(10)}

\author{{\bf K.S.Babu and S.M.Barr}\\
Bartol Research Institute\\ University of Delaware\\
Newark, DE 19716}

\date{BA-94-04}
\maketitle

\begin{abstract}
It is shown that a natural gauge hierarchy and doublet-triplet
splitting can be achieved in SO(10) using the Dimopoulos-Wilczek
mechanism. Artificial cancellations (fine-tuning) and arbitrary
forms of the superpotential are avoided, the superpotential
being the most general compatible with a symmetry. It is shown by example
that the Dimopoulos-Wilczek mechanism can be protected against the
effects of higher-dimension operators possibly induced by Planck-scale
physics. Natural implementation of the mechanism
leads to an automatic Peccei-Quinn symmetry. The same local symmetries
that would protect the gauge hierarchy against Planck-scale effects
tend to protect the axion also. It is shown how realistic quark and
lepton masses might arise in this framework. It is also argued
that ``weak suppression'' of proton decay
can be implemented more economically than can ``strong suppression'',
offering some grounds to hope (in the context of SO(10)) that
proton decay could be seen at Superkamiokande.
\end{abstract}
\newpage

\dspace

\section{Introduction}

In this paper we present a realistic and natural supersymmetric
grand-unified theory (SUSY-GUT) based on the gauge group SO(10).
By a natural SUSY-GUT we mean one that satisfies the following three
conditions. (1) The doublet-triplet splitting (or ``2/3 splitting'')
of the Higgs multiplet does not involve artificial cancellations or
``fine-tuning'' of parameters; (2) the superpotential has the most
general form allowed by some symmetry principle; and (3) local
symmetry prevents the appearance in the effective sub-Planck-scale
theory of higher dimension operators that would disrupt the gauge
hierarchy.

In a previous paper$^{(1)}$ we explored an elegant mechanism, proposed
originally in 1981 by Dimopoulos and Wilczek,$^{(2)}$ which achieves
the 2/3 splitting without artificial cancellations or fine tuning.
There we showed that this mechanism makes possible a simple suppression
of proton decay coming from the dimension-five operators mediated by
the exchange of the color-triplet higgsinos, while at the same time
preserving the wonderfully successful SUSY-GUT prediction$^{(3)}$ of
$\sin^2\theta_W$.

The Dimopoulos-Wilczek (DW) mechanism, which accomplishes all this,
calls for the group SO(10) rather than SU(5) or flipped SU(5)$^{(4)}$
$[SU(5) \times U(1)]$. SO(10) has long been regarded as an especially
attractive
gauge group for grand unification$^{(5)}$ for a number of other reasons as
well. (a) It unifies a whole family into a single irreducible representation.
(b) It is automatically anomaly free. (c) It completes each family with a
right-handed neutrino. And (d), it automatically builds in ``matter parity'',
since quarks and leptons are in spinors and Higgs fields are in tensors
of SO(10). [Matter parity is a $Z_2$ subgroup of the $Z_4$ center of
SO(10), whereas the centers of SU(5) and $E_6$ do not contain matter
parity.]

In Reference 1 it was not shown that the DW mechanism is natural in the
strong sense which is used in this paper and which is defined above.
While it was shown that 2/3 splitting is achieved without artificial
cancellations (condition 1), the superpotentials studied in Ref. 1
were \underline{not} the most general allowed by some symmetry
(condition 2); rather, certain terms were simply left out. This is
certainly natural in the weaker sense that the non-renormalization theorems
of supersymmetry would prevent such omitted terms from being induced by
radiative corrections, but it is nevertheless arbitrary.
In section 2 we present an SO(10)
model which satisfies the first two naturalness conditions. This model
is quite simple (simpler than those studied in Ref. 1) and perhaps
even a minimal SO(10) SUSY-GUT.

We treat the problem of studying the third naturalness condition in section
3. The reason we treat this separately is that it is unknown whether
Planck-scale physics does in fact induce all possible higher-dimension
operators (not forbidden by local symmetry) in the effective sub-Planck-
scale lagrangian,$^{(6)}$ and, if so, how large their coefficients might be.
Conceivably there could be some tremendous suppression that would make it
unnecessary to worry about these effects at all. In section 3, however,
we will make the most pessimistic assumption that such operators are
suppressed only by the dimensionally appropriate powers of the Planck
mass: that is, that they are as large as they can be. Even under this
assumption it is shown that a straightforward extension of the model
of section 2 which is only slightly more complicated can prevent any
disruption of the gauge hierarchy by Planck-scale physics.

In the models of both sections 2 and 3 the symmetries imposed to make
the gauge hierarchy natural are closely akin to Peccei-Quinn
symmetries$^{(7)}$ and in fact lead automatically to the existence of
an invisible axion. This is not completely surprising, since the
terms that have to be prevented are those that would produce a large
$\mu$-parameter, and it is well-known that a $\mu$-parameter can be
prevented by a Peccei-Quinn symmetry.$^{(8)}$ However, we find it
significant that a natural gauge hierarchy in SO(10) may not only
require the existence of an axion but that in fact the local symmetry
that may be needed to protect the hierarchy from Planck-scale-induced
higher-dimensional operators may protect the axion from such effects
as well.$^{(9)}$ This will be discussed in section 4.

Finally, there is the question of realistic quark and lepton masses.
The question is how to avoid the ``bad'' predictions of SO(10), namely
that the up and down quark mass matrices are proportional $(m_t^0:
m_c^0:m_u^0 = m_b^0:m_s^0:m_d^0)$ and that $m_{\mu}^0 = m_s^0$ and
$m_e^0 = m_d^0$. The prediction $m_{\tau}^0 = m_b^0$ is, of course,
the ``good'' one; and so the obvious and common suggestion is that
higher order effects disturb the relations involving the lighter
generations. It turns out that this suggestion is easy to implement
in the context of the ideas discussed here. In fact these ideas lend
themselves very well to a promising approach to the quark and lepton mass
puzzle that has already been advocated in the literature.$^{(10)}$
This will be discussed in section 5.
In section 6 we will summarize our results and conclusions.

\section{A Simple SO(10) SUSY-GUT}

Consider a supersymmetric SO(10) model with chiral superfields
in the following representations: three {\bf 16}'s of quarks and
leptons $(F_I, I=1,2,3)$, and Higgs fields that are in two
{\bf 10}'s $(T_1, T_2)$, three {\bf 45}'s $(A,A',A^{\prime\prime})$, a
{\bf 54} $(S)$,
a $\overline{{\bf 126}}$ $(\bar{C})$ and a {\bf 126} $(C)$. Let there also be
a $Z_3$ discrete symmetry under which the fields transform as in Table I.
Then the most general SO(10)$\times Z_3$-invariant, renormalizable
superpotential is given by

\begin{equation}
\begin{array}{ccl}W & = & \sum_i W_i\strut\\
W_1 & = & \sum_{I,J} \lambda_{IJ} F_I F_J T_1 + \sum_{I,J}
\lambda_{IJ}^{\prime} F_I F_J \bar{C}\strut\\
W_2 & = & \lambda_1 T_1 A T_2 + M_2(T_2)^2 + \lambda_2 T_2 S T_2\strut\\
W_3 & = & M_A A A' + \lambda_3 S A A' + M_S S^2 + \lambda_4 S^3\strut\\
W_4 & = & M_A^{\prime\prime} A^{\prime\prime 2} + \lambda_5
S A^{\prime\prime 2} + M_C \bar{C}C + \lambda_6
\bar{C}A^{\prime\prime}C\strut\\W_5 & = & \lambda_7AA'A^{\prime\prime}.
\end{array}
\end{equation}

\noindent
We assume that the dimensionless couplings, $\lambda_i$, that appear
in W are of order one and that the mass parameters are of order
$M_{GUT} \simeq 10^{16}GeV$. (As will be seen shortly, however,
$M_2 + \lambda_2\left< S\right>$ must be slightly less than $M_{GUT}$
-- about $10^{15}GeV$ -- in order to suppress proton decay.) Consequently,
the VEVs of $A,A',A^{\prime\prime},S,\bar{C}$, and $C$ will be of
order $10^{16}GeV$.

The function of each term in $W$ is easily understood. $W_1$ is to give
mass to the quarks and leptons: $FFT_1$ gives Dirac mass to the light
fermions and $FF\bar{C}$ gives Majorana masses to the right-handed
neutrinos. (Of course, by themselves the terms displayed in $W_1$ would
not give a realistic pattern of masses, as noted in the introduction.
Some attractive and economical ideas for improving this situation are
considered in section 5.)

$T_1$ contains the light Higgs doublets, $H$ and $H'$, of the supersymmetric
standard model as well as their color-triplet partners, $H_3$ and $H_3'$,
that must be made superheavy. This is the well-known 2/3 splitting
referred to in the introduction, which here is accomplished by the
Dimopoulos-Wilczek sector, $W_2$. The VEV of the adjoint, $A$, is assumed to
have the DW form, $\left<A\right> = diag(a,a,a,0,0)\times\left( \begin{array}
{cc} 0 & 1\\-1 & 0\end{array}\right)$. Thus the color-triplets in $T_1$ and
$T_2$ have a mass matrix of the form

\begin{equation}
\left( \bar{3}(T_1)\;\bar{3}(T_2)\right) \left( \begin{array}{cc}
0 & \lambda_1 a\\-\lambda_1 a & M_2 + \lambda_2\left<S\right>\end{array}
\right) \left( \begin{array}{c} 3(T_1)\\3(T_2) \end{array} \right),
\end{equation}

\noindent
which makes all of them superheavy, as $M_2$, $a$, and $\left<S\right>$
are all of or near the GUT scale, $M_{GUT} \simeq 10^{16}GeV$. The doublets,
on the other hand, have a mass matrix (ignoring weak-scale effects) of
the form

\begin{equation}
\left( \bar{2}(T_1) \; \bar{2}(T_2) \right) \left( \begin{array}{cc}
0 & 0\\0 & M_2 + \lambda_2\left<S\right>\end{array} \right) \left(
\begin{array}{c} 2(T_1)\\2(T_2) \end{array} \right),
\end{equation}

\noindent
so that the $2\;(\equiv H)$ and the $\bar{2}\;(\equiv H')$
in $T_1$ are light, as required. Higgsino-mediated proton decay happens
through the diagram in Fig. 1.  This gives a proton-decay amplitude
proportional to $(M_2 + \lambda_2\left<S\right>)/(\lambda_1a)^2$, as
can be seen directly by inverting the mass matrix in Eq. (2). Thus the
proton-decay rate from colored-higgsino exchange has a suppression
factor $[(M_2 + \lambda_2 \left< S \right>)/ \lambda_1a]^2$ which allows
a comfortable consistency with present experimental limits if
$M_2 + \lambda_2 \left< S \right>$ is about $10^{15}GeV = 10^{-1}M_{GUT}$.
This is what is called ``weak suppression'' of proton decay in Ref. 1.

[The color-triplet higgsinos in the $\overline{{\bf 126}}$ multiplet
can, in
general, also induce proton decay, with an amplitude proportional to the
neutrino Majorana mass matrix. However, this does not happen in the present
model, since $Z_3$ prevents terms such as $\bar{C}^2S$ and since the
color-triplet in $\bar{C}$ does not mix with the color-triplets in $T_1$
or $T_2$.]

The purpose of $W_3$ is to give the required DW form to $\left< A \right>$
and $\left< A'\right>$. If $\left<S\right> \equiv diag(s,s,s,-\frac{3}{2}s,
-\frac{3}{2}s)\times\left(\begin{array}{cc}1 & 0\\0 & 1\end{array}\right)$
and $\left<A\right> \equiv diag(a,a,a,b,b) \times \left(\begin{array}{cc}
0 & 1\\ -1 & 0\end{array} \right)$ then $F_{A^{\prime}} = 0$ gives

\begin{equation}
\begin{array}{ccc}(M_A + \lambda_3s)\;a & = & 0,\strut\\(M_A - \frac{3}{2}
\lambda_3s)\;b & = & 0.\end{array}
\end{equation}

\noindent
One solution is $b=0,a\neq 0,$ and $s= -M_A/\lambda_3$. The $F_A=0$
equation then forces $b' = 0$ where $\left<A'\right> \equiv diag (a',a',
a', b',b') \times \left(\begin{array}{cc} 0 & 1 \\ -1 & 0\end{array}
\right)$ so that $\left< A'\right>$ also has
the DW form. The magnitude of the combination $aa'$ is fixed by the
$F_S=0$ equation, but the individual magnitudes of $a$ and $a'$ are
not determined until SUSY is broken, when the soft terms
$\mid A\mid ^2$ and $\mid A'\mid ^2$, whose coefficients are both of order
$M_{SUSY}^2\stackrel{_<}{\sim} (1TeV)^2$, will set $a\sim a'\sim M_{GUT}$.

The form of $W_3$ differs in a very significant way from the form considered
by Srednicki in Ref. 11 and adopted in Ref. 1. The Srednicki form is
$S^3 + M_SS^2 +SA^2 + M_AA^2$. Here $A^2$ has been replaced by $AA'$.
The reason for this is the necessity of ruling out an $M_{GUT}(T_1)^2$
term, which would destroy the gauge hierarchy by making all the Higgs
doublets superheavy. (This is just a $\mu$-term with $\mu$ of order
$M_{GUT}$.) The connection with the form of $W_3$ is seen through
Fig. 2. The point of Fig. 2 is not that this diagram itself is large --
in fact, due to the non-renormalization theorems of SUSY, it will be
suppressed by the SUSY-breaking scale and not endanger the gauge hierarchy.
Rather, the point is that if such a diagram can be drawn it implies that
no symmetry of the theory forbids a $(T_1)^2$ term, and therefore,
notwithstanding the non-renormalization theorems that allow this term
to be omitted, it is unnatural in the sense of our second condition to
leave it out. The vertex $T_1AT_2$ that appears in Fig. 2 is a necessary
ingredient of the DW mechanism, and the term $M_2(T_2)^2$ is required
if there are not to be {\bf two} pairs of light Higgs doublets,
which would be very bad for $\sin ^2\theta_W$.$^{(1)}$ Thus any hope of
preventing the $(T_1)^2$ term by symmetry requires the vertex $MA^2$
which appears in Fig. 2 not to exist. Hence our replacement of it by
$MAA'$.

The effect of having only the combination $AA'$ appear, required to
forbid $(T_1)^2$, is that the model has an ``accidental'' Peccei-Quinn
symmetry under which $A \rightarrow e^{i \alpha} A, A' \rightarrow e^{-i
\alpha} A', T_1 \rightarrow e^{-i \alpha} T_1$, and $F_I \rightarrow
e^{i \alpha/2}F_I$. The resulting axion is invisible, as $f_a = \left<
A \right> \sim M_{GUT}$. This will be discussed further in section 4.

The $W_4$ sector generates SU(5)-singlet VEVs for $A^{\prime\prime},
\bar{C}$, and $C$ that break the rank of SO(10) down to that of the
standard model and gives right-handed-neutrino masses. $W_3$ and $W_4$
together break SO(10) $\rightarrow SU(3) \times SU(2) \times U(1)$.

As explained in Ref. 1, the term $\lambda_7AA'A^{\prime\prime}$ couples
the $(A,A',S)$ and $(A^{\prime\prime}, \bar{C},C)$ sectors together
in such a way as to prevent goldstone bosons without destabilizing the
DW form of $\left< A \right>$ and $\left< A' \right>$.

The model given in Eq. (1) is close to being a minimal realistic and natural
SO(10) SUSY-GUT. It breaks SO(10) completely to the standard model, avoids
extra light fields (goldstone or otherwise) that would disturb $\sin ^2
\theta_W$, gives natural 2/3 splitting, allows higgsino-mediated proton
decay to be suppressed, and produces quark and lepton masses and the
see-saw mechanism for neutrino masses. It is hard to imagine doing all
this with a smaller set of fields or fewer couplings. At least one
{\bf 45} of Higgs fields is needed to break SO(10) and do 2/3 splitting.
Because of the antisymmetry of {\bf 45} one needs {\bf two}
{\bf 10}'s to write the term ${\bf 10}_1 {\bf 45}\; {\bf 10}_2$.
(As emphasized in Ref. 1 such a doubling is probably necessary in any
case to suppress proton decay.) Getting the DW form of $\left< {\bf 45}
\right>$ is done most simply with an auxiliary {\bf 54}. Breaking
the rank of SO(10) requires at least either $\overline{{\bf 126}} +
{\bf 126}$ or $\overline{{\bf 16}} + {\bf 16}$. And the tripling of
adjoint representations serves two important purposes: allowing the
$AA'A^{\prime\prime}$ term that couples different sectors together
without destabilizing VEVs and allowing a Peccei-Quinn-type symmetry
to keep the doublets $H$ and $H'$ light.

Certainly this model is not unique. For example, there are other ways to
achieve the DW form of VEVs or to break SO(10). But other choices do
not seem to lead to models that are simpler with respect to the number of
fields or the number of terms in $W$.

{}From an experimental point of view a very important question is
whether higgsino-mediated proton decay can be ``strongly suppressed''
(in the terminology of Ref. 1)
in a fully natural way. The answer is yes, but it appears that
a model of considerably greater complexity is required. We have already
had to use three adjoints to get VEVs in the DW form. (Note that
because we were forced to have terms $AA'$ and $SAA'$ to prevent
$(T_1)^2$, the VEVs of $A$ and $A'$ are both compelled to have the DW
form.) To get a VEV in the upside-down DW form, $diag(0,0,0,b,b)
\times \left( \begin{array}{cc} 0 & 1\\ -1 & 0 \end{array} \right)$,
required for ``strong suppression'' of proton decay would involve at
least two {\bf more} adjoints for a total of five.

On the grounds of economy it is justified to say that in SO(10)
proton decay is more likely to be suppressed ``weakly'' (ie. numerically
by factors of order one) than ``strongly''. This holds out some hope
that proton decay mediated by higgsinos can be seen experimentally.

\section{The Problem of Higher-Dimension Operators}

As noted in the introduction, it is not known whether Planck-scale physics
necessarily induces all possible higher dimension operators allowed by
local symmetry into the effective theory below the Planck scale,$^{(6)}$
and if they do how large the coefficients of such operators would be. If such
effects are negligible then the relatively simple SUSY-GUT presented
in the previous section is adequate (except for realistic quark and
lepton masses, which will be dealt with in section 5).

In this section we will assume the ``worst'', namely that every higher-
dimension operator allowed by local symmetry is present suppressed only
by dimensionally appropriate powers of the Planck mass. We will show
by explicit example that the Dimopoulos-Wilczek mechanism can give
a natural gauge hierarchy even under this assumption. This will involve
identifying the dangerous higher-dimensional operators, exhibiting a
local symmetry that rules them out, and showing that this symmetry can
satisfy anomaly constraints. The symmetry used in the model constructed
in this section is a single local U(1) for simplicity of analysis.
A combination of symmetries, either continuous or discrete, could
also play the same role.

There are two kinds of higher-dimensional operators that endanger the gauge
hierarchy, those that contain $(T_1)^2$ or $T_1\cdot T_2$ and directly give
a mass term to the light doublets, and those that destabilize the DW form
of the VEV of $A$. In a model with only two {\bf 10}'s of Higgs fields,
such as that in section 2, it is very difficult to prevent the first kind
of operator by symmetry. The reason can be seen by considering Fig. 3.
This diagram uses vertices $T_1AT_2$ and $M(T_2)^2$ that must exist
in a satisfactory model with only two {\bf 10}'s; the first for the
DW mechanism, and the second to avoid an extra pair of light doublets.
The resulting operator is not itself dangerous, as it involves the
contraction $T_1\cdot A\cdot A\cdot T_1$ and thus, because of the DW
form of $\left< A \right>$, does not give mass to the doublets in $T_1$.
However, if this term were allowed by local symmetry, then so also
would be $(T_1\cdot T_1)$tr$(A\cdot A)/M_{Pl}$ which, by assumption,
would be induced by Planck-scale physics and would give a superlarge mass
of order $\mu \sim M_{GUT}^2/M_{Pl}$ to the doublets in $T_1$. This problem
could be alleviated somewhat if the mass of $T_2$ came from a cubic term,
$N(T_2)^2$ where $N$ is a singlet superfield,
instead of an explicit mass term, $M(T_2)^2$. Then if the
superpotential contained $N^3$ or $MN\bar{N}$ a local symmetry would
allow $(T_1\cdot T_1)$tr$(A\cdot A)N^2/M_{Pl}^3$ or
$(T_1\cdot T_1)$tr$(A\cdot A)\bar{N}/M_{Pl}^2$ giving $\mu$ to be of
order $M_{GUT}^4/M_{Pl}^3 \sim 10^7 GeV$ or $M_{GUT}^3/M_{Pl}^2 \sim
10^{10} GeV$, respectively; in either case too large.

One could imagine an R-symmetry that would prevent such terms. For
example, consider $R \times Z_3$, where under $R$ all superfields
transform as $\phi \rightarrow -\phi$ and $W \rightarrow -W$, and under
$Z_3$ all superfields transform as $\phi \rightarrow e^{2 \pi i/3}\phi$
and $W\rightarrow W$. Then $N(T_2)^2$ and $T_1AT_2$ would be allowed
but no higher-dimension term would be allowed until d = 9. However, we
have found no satisfactory model where the $T_1^2A^2$ terms are
forbidden using R-symmetries.

Another approach which we have found to lead to a fully realistic
scenario involves the existence of three {\bf 10}'s, which
will be denoted $T_1$, $T_2$, and $T_3$. Suppose there are terms
$T_1AT_2$, $QT_2T_3$, and $P(T_3)^2$, where $P$ and $Q$ denote singlet
superfields with VEVs $\sim M_{GUT}$ (later, for the sake of economy, we will
identify $Q$ with $\bar{P}$ in our illustrative model),
and $\left< A \right>$ has the
DW form. Then the $3\times 3$ mass matrix for the color-triplets
in $T_i$ has rank 3, while the mass matrix for the doublets has rank
2, with the massless doublet being in $T_1$. The point of this
straightforward generalization of the model of section 2 is that the analogue
of the diagram of Fig. 3, which is shown in Fig. 4, gives an operator of
at least dimension 7: $(T_1\cdot A\cdot A\cdot T_1)\bar{Q}^2P$. (It is
assumed there is a field $\bar{Q}$ that has coupling $\bar{Q}Q$. If there is no
such field then Fig. 4 will give an operator of dimension even greater than
7.) Again, it is not this operator itself which of interest, for given
the DW form of $\left< A \right>$ and the way it is contracted, this
operator contributes nothing to $\mu$. But if such a term is allowed then
so is $(T_1)^2$tr$(A)^2\; \bar{Q}^2P/M_{Pl}^4$, which gives $\mu \sim
M_{GUT}^5/M_{Pl}^4 \sim 10\;TeV$. Given that this is a very crude dimensional
estimate, this is satisfactorily close to the weak scale.

Contributions to $(T_1 \cdot T_2)$, being off-diagonal in the mass matrix
(see Eq. (3)), need only be suppressed to $O(1/M_{Pl}^2)$ to prevent $\mu$
from being larger than $O(1/M_{Pl}^4)$. Thus dimension-four operators
like $T_1 \cdot T_2$tr$(A\cdot A^{\prime\prime})$ must be forbidden
by local symmetry. Given that the term $T_1AT_2$ must exist (for the
DW mechanism), it must be that $A^{\prime\prime}$ transforms non-trivially
under the local symmetry.

Finally, there are the operators that destabilize the DW form of
$\left< A \right>$. If $\left< A \right> \sim M_{GUT} \cdot diag(1,1,1,
\epsilon ,\epsilon )\times \left( \begin{array}{cc} 0 & 1\\ -1 & 0
\end{array} \right)$, then, since $A$ appears off-diagonally in the
mass matrix of the $T_i$, $\mu (T_1)^2$ is induced with $\mu \sim \epsilon
^2 M_{GUT}$. Thus $\epsilon$ must be $\stackrel{_<}{_\sim} (M_{GUT}/M_{Pl})^2$,
and the destabilizing term for $\left< A \right>$ must be at least fifth
order in superfields.

A dangerous term would be $(A\cdot A' \cdot \bar{C} \cdot C)$.
{}From the fact that under $SU(2)_L \times SU(2)_R \times SU(4)_C \subset
SO(10)\; \; \; \; \; \left< \bar{C} \right> \sim (1,3,10), \left< C \right>
\sim (1,3,\overline{10})$, and $\left< A' \right> \sim (1,1,15)$, it
follows that this term acts as a linear term for the $(1,3,1)$ component
of $A$. (The DW form is in (1,1,15).)

Armed with this information, one can write down an $SO(10) \times U(1)$
model with the fields given in Table II, that avoids all higher-dimension
operators dangerous to the gauge hierarchy. The form of the superpotential
(including certain relevant higher-dimension operators) is

\begin{equation}
\begin{array}{ccl} W & = & F_IF_JT_1 + F_IF_J\bar{C}\strut\\
& + & T_1AT_2 + \bar{P}T_2T_3 + P(T_3)^2\strut\\
& + & M_{GUT}AA' + SAA' + M_{GUT}S^2 + S^3\strut\\
& + & \bar{P} \bar{C} C + \bar{C} A^{\prime\prime} C + (A^{\prime\prime})^2
(P^2/M_{Pl})\strut\\
& + & A A' A^{\prime\prime}(P/M_{Pl})\strut\\
& + & m_P \bar{P} P + (\bar{P} P)^2/M_{Pl},
\end{array}
\end{equation}

\noindent
where each term has a dimensionless coefficient which has not been
written.

The structure of Eq. (5) is easily understood by comparison with Eq. (1).
Since $A^{\prime\prime}$ has to transform non-trivially (as noted above,
to forbid $T_1 \cdot T_2$tr$(A \cdot A^{\prime\prime})$) the terms
$AA'A^{\prime\prime}$ and $MA^{\prime\prime 2}$ have now to come
from higher-dimension operators (which, by assumption, are present).
Similarly, because the singlet, $P$, has a non-trivial U(1) charge,
higher-dimension operators must be taken into account if it is to get
a VEV. Since $\left< P \right> \sim \left< \bar{P} \right> \sim
(m_P M_{Pl})^{\frac{1}{2}}$, it must be that $m_P$ is $\sim M_{GUT}^2/
M_{Pl}$. this in turn implies that $\left< \bar{C} \right> \sim \left<
C \right> \sim M_{GUT}^{\frac{3}{2}} M_{Pl}^{-\frac{1}{2}} \sim M_{GUT}/30$.
[It should be noted that the soft SUSY-breaking terms, $\mid \bar{C}
\mid ^2$, $\mid C \mid ^2$, etc., will insure that $\left< \bar{C}
\right> \sim \left< C \right>$ and $\left< \bar{P} \right> \sim
\left< P \right>$.]

There are three anomaly conditions to be satisfied: $SO(10)\times U(1)$,
gravity$\times U(1)$, and $U(1)^3$. Since there are only two unknowns,
p and q (see Table II), these equations are over-determined. But it
is clear that additional gauge singlets will contribute to the
gravity$\times U(1)$ and $U(1)^3$ anomalies without in any way affecting
the issues discussed up to this point. For example, fields $N_i +
\bar{N}_i$ with charges $n_i$ and $(-n_i + p)$ can get mass from
$\bar{N} N \bar{P}$ and contribute $+p$ to the gravity$\times U(1)$ anomaly
and $3n_i^2 p -3 n_i p^2 + p^3$ to the $U(1)^3$ anomaly while leaving
the $SO(10)^2 \times U(1)$ condition unaffected.  The addition
of such fields allows all the anomaly conditions to be satisfied. There
could, of course, be also additional SO(10)-non-singlet fields.

\section{Peccei-Quinn Symmetries and Axions}

It is apparent that the model exhibited in the last section, like
that of section 2, has an ``accidental'' Peccei-Quinn symmetry under
which $A \rightarrow e^{i \alpha} A,\; A' \rightarrow e^{-i \alpha} A',
\; T_1 \rightarrow e^{-i \alpha} T_1,$ and $F_I \rightarrow e^{i \alpha
/2} F_I$. In fact, every vector (p,q), where p and q are defined by
Table II, corresponds to the generator of a U(1), which we can denote
by $U(1)_{(p,q)}$, so that there are two linearly
independent U(1) symmetries. One linear combination is anomaly free,
by construction, and is local. The other U(1) has an $SO(10)^2 \times U(1)$
(and thus an $SU(3)_C^2\times U(1)$) anomaly and qualifies as a Peccei-
Quinn symmetry.

The $SO(10)^2 \times U(1)$ anomaly-cancellation condition can be written
as $mp + nq = 0$, where m and n are integers that depend on the particle
content of the model. Then the local U(1) is just $U(1)_{(n,-m)}$.
Each operator will have a charge under $U(1)_{(p,q)}$ of $Mp+Nq$ where
M and N depend on the charges of the fields of which it is composed.
If $M=N=0$ then the operator trivially is invariant under the full
$U(1) \times U(1)$. If $(M,N) \propto (m,n)$ then the operator will be
invariant
under the local U(1) but not under the global U(1), that is, the Peccei-
Quinn symmetry. Under the assumption of the previous section such an
operator will be induced by quantum gravity in the effective superpotential
below the Planck scale, and, if it has dimension D and is composed entirely
of fields that have VEVs of order $M_{GUT}$, then it will contribute
$\Delta m_a \sim M_{GUT}^{D-3}/M_{Pl}^{D-4}$ to the axion mass. To
solve the strong CP problem by the Peccei-Quinn mechanism this must
be less than the QCD-instanton-generated axion mass, which in this
case is given by $m_a^{QCD} \sim \Lambda_{QCD}^2/f_a \sim \Lambda_{QCD}
^2/M_{GUT}$, and thus D must be greater than 14. In the model constructed
in the previous section, the lowest-dimension operator that contributes
to the axion mass will be of dimension $D= m+n$ where m and n are
defined as above and are normalized to be relatively prime integers.

To take a concrete example, suppose that the quarks and leptons, $F_I$,
of the model of section 3 have charge $\frac{1}{2} (\frac{3}{2} p +
q)$ so that they get mass directly from the term $\sum_{I,J}
\lambda_{IJ} F_I F_J T_1$, and that there are no other SO(10)-non-
singlet fields in the theory except those listed in Table II. Then
the $SO(10)^2 \times U(1)$ anomaly-cancellation condition turns out
to be $31p +2q = 0$. Consequently, the lowest dimension operators
respecting $U(1)_{local}$ that break $U(1)_{PQ}$ are of dimension
33. For example, one such is $A'^2\bar{P}^{31}$. (See Table II.)

It is apparent that the charge assignments and thus the dimension of
the smallest PQ-violating operators are dependent upon details of
the model. To take another example, if the quark and lepton masses
came instead from a higher-dimension operator like $F_IF_JT_1A^{\prime
\prime}/M_{GUT}$ then the charge of $F_I$ would be $\frac{1}{2}
(\frac{5}{2} p + q)$, the anomaly condition would be $17p + q =0$,
and the lowest-dimension PQ-violating operator would have D=18.

Charges such as these seem rather bizarre. But this model, wherein
the local symmetry that controls the Planck-scale physics is just a single
U(1), is presented merely as an illustration. The true theory could
have, instead of a single U(1), a product of several discrete or continuous
symmetries. Then even with the fields having smaller charges the smallest
operator that respects all the local symmetries but violates
$U(1)_{PQ}$ could be of high dimension as required.

The important lesson that the illustrative examples teach is that the
same local symmetry that protects the gauge hierarchy from the (possible)
effects of quantum gravity will also tend to preserve the axion solution
to the strong CP problem. It should be emphasized that the axion decay
constant, $f_a$, is of order $M_{GUT} \sim 10^{16} GeV$, and thus there
is a potential ``axion energy problem'' for cosmology.$^{(12)}$ However,
in inflationary scenarios this problem is not necessarily real.$^{(13)}$
But laboratory searches for axionic dark matter would have a difficulty
due to the small cross-sections.

\section{Realistic Quark and Lepton Masses}

The simple Yukawa term that appears in eq.(1), $\sum_{I,J} \lambda_{IJ}
F_IF_JT_1$, gives the relations at the GUT scale $M_{lepton} =
M_{down \; quark} \propto M_{up \; quark}$, and is therefore not satisfactory.
In
order that these characteristic SU(5) and SO(10) predictions be modified
it is necessary that the light fermion mass matrices feel the effects
of the breaking of the grand-unified symmetries. The simplest possibility
is that higher-dimensional operators contribute to them; for instance,
$F_IF_JT_1 \tilde{A}/M$, where $\tilde{A}$ is some SO(10)-non-singlet
field that has a VEV of order $M_{GUT}$.
If $M=M_{Pl}$ this term yields contributions to the
quark and lepton mass matrices which are too small
to cure the problem. (We are assuming that $\tan \beta \cong m_t/m_b$,
which would be typical of a simple SO(10) model, in which case the
corrections to the lepton and down-quark matrices from gravity would
be at most of order $10^{-3}m_{\tau}$ and $10^{-3}m_b$.) Thus it would
appear that M must be of order $M_{GUT}$, which implies that these
operators are induced not by Planck-scale physics but by integrating
out fields with mass $O(M_{GUT})$ appearing in the effective sub-Planck-
scale theory. In other words, new fields must be introduced into the
theory.

A most economical possibility is to introduce a vectorlike pair
consisting of a family and an anti-family, $F + \bar{F} = {\bf 16}
+ \overline{{\bf 16}}$. Then diagrams like that shown in Fig. 5 become
possible. (In that figure, $M$ could be replaced by the VEV of a
Higgs field in the product ${\bf 16} \times \overline{{\bf 16}}
= {\bf 1 + 45 + 210}$, and/or $\tilde{A}$ could be replaced by a
singlet, depending on the model.) What
makes this such an economical and compelling idea is that it not only
allows SU(5) and SO(10)-breaking effects to be introduced into the
light quark and lepton mass matrices, but it also suggests a qualitative
explanation for various observed features of those masses. In fact
the kind of diagram shown in Fig. 5 is the basis of a model of light-fermion
masses that has already appeared in the literature.$^{(10)}$

To understand this idea in more detail, consider adding to the model
of section 3 the fields $F + \bar{F}$. And imagine that there is an adjoint
Higgs field, $\tilde{A}$, whose VEV points in a direction Q which
is a non-trivial linear combination of B-L and Y (weak hypercharge).
(Q is, of course, a generator of SO(10).) Let $F$, $\bar{F}$, $F_I$,
and $\tilde{A}$ be assigned $U(1)_{local}$ charges such that the diagram
in Fig. 5 exists, but that the cubic term $F_IF_JT_1$ is forbidden.
Then the effective mass term that arises from Fig. 5 is

\begin{equation}
\kappa \sum_{I,J} f^c (F_I) \{ \hat{b} _I \hat{a} _J \vec{Q} +
 \bar{Q} \hat{a} _I
\hat{b}_J \} f(F_J) \cdot \left<T_1 \right>.
\end{equation}

\noindent
The Yukawa couplings, $a_I$ and $b_I$, as shown in Fig. 5, are those
that couple $F_I$ to $\bar{F}$ and $F$: $\sum_I a_I F_I \bar{F} \tilde{A}
+ \sum_I F_I F T_1$. And the hatted quantities in Eq. (6) are simply
defined by $ \hat{a}_I \equiv a_I / \mid a \mid $
and $\hat{b}_I \equiv b_I / \mid b \mid $. The magnitudes of the couplings
and the superheavy VEVs that appear in Fig. 5 are all combined in the
factor $\kappa$. $f^c(F_I)$ and $f(F_I)$ are the anti-fermion and
fermion that are contained in $F_I$. For the down quark matrix,
for instance, $f^c(F_I) = d^c_I$ and $f(F_I) = d_I$.
Note that the Yukawa couplings, $a_I$ and $b_I$, are vectors in ``family
space'' rather than $3 \times 3$ matrices as is the $\lambda _{IJ}$
of Eq. (1). Thus a ``factorized'' form of the mass matrices results.$^{(14)}$
If there were only one such factorized term the mass matrices would
be rank-1, but as there are two terms in Eq. (6) each mass matrix is
rank-2, thus explaining the extreme lightness of the first generation
(which, of course, is assumed to get mass from some other, perhaps higher
order, operator.)

One may, without loss of generality, choose the basis in family space
so that $\hat{a}_I = (0,0,1)$ and $\hat{b}_I = (0,\sin \theta,
\cos \theta )$. Then for the $3 \times 3$ mass matrix of fermions of type
$f$ ($f = \ell ^{-}$, d, or u) one has simply

\begin{equation}
M^{(f)}=\kappa \cdot v^{(f)} \cdot \left( \begin{array}{ccc}0 & 0 & 0\\
0 & 0 & Q_f \sin \theta\\0 & Q_{f^c} \sin \theta &
(Q_f + Q_{f^c}) \cos \theta \end{array} \right)
\end{equation}

\noindent
where $v^{(f)}$ is defined to be $v$ for $f=u$ and $v'$ for $f=d$ or
$\ell^-$. Aside from the fact that they have rank 2 (thus explaining
the lightness of $e^-$, $u$, and $d$) these matrices have other
significant features. If $\sin \theta$ is somewhat small, then the
second generation has mass of order $\frac{1}{4} \tan ^2 \theta$ times
that of the third generation. Thus $\sin \theta \sim \frac{1}{3}$
would be sufficient to explain the hierarchy between the two heavier
generations. But most interesting is that the form of these matrices
explains why $m_b^0 \cong m_{\tau}^0$ while $m_{\mu}^0/m_{\tau}^0$,
$m_s^0/m_b^0$, and $m_c^0/m_t^0$ are all different. For, neglecting
effects of order $\sin ^2 \theta$,

\begin{equation}
\frac{m_b^0}{m_{\tau}^0} = \frac{Q_d + Q_{d^c}}{Q_{\ell ^-} +
 Q_{\ell ^+}}=1,
\end{equation}

while

\begin{equation}
\frac{m_s^0}{m_{\mu}^0} \cong \frac{Q_d Q_{d^c}/(Q_d+Q_{d^c})}
{Q_{\ell ^-} Q_{\ell ^+} /(Q_{\ell ^-} + Q_{\ell ^+} )} =
\frac{Q_d Q_{d^c}}{Q_{\ell ^-} Q_{\ell ^+} } \neq 1.
\end{equation}

The second equality in Eq. (8) follows from the fact that the same
Higgs field, $H'$, couples to $d^c d$ and to $\ell ^+ \ell ^-$,
and thus $Q_d + Q_{d^c} = Q_{\ell ^-} + Q_{\ell ^+} = - Q(H')$ for
{\bf any} generator $Q$; whereas $Q_dQ_{d^c} \neq Q_{\ell ^-}
Q_{\ell ^+}$ in general.

We see that the simple diagram of Fig. 5 can explain why $m_b^0 =
m_{\tau}^0$ is the good relation, while the other SU(5) and SO(10)
relations are not good, why the first generation is extremely
light compared to the second and third, and can give ``Fritzschian'' relations
between the mass ratios and mixing angles. But obviously this is
not the whole story, for some other diagram is needed to give non-zero
masses to the first generation. Also, the field denoted $\tilde{A}$
cannot be one of the adjoints $A$, $A'$, or $A^{\prime\prime}$ of
the model of section 3, for $Q$ cannot be either purely $B-L$ or
purely $X$ ($X$ being the generator of the U(1) in $SU(5) \times U(1)
\subset SO(10)$). For $Q=B-L$ would give a vanishing (3,3) element in
Eq. (7),
and $Q=X$ would give $m_s^0 \cong m_{\mu}^0$. Nevertheless, the basic
approach proposed in Ref. 10 and reviewed here clearly has many
attractive features and
seems quite compatible with the Dimopoulos-Wilczek framework that
we have elaborated in the previous sections. It should be emphasized
that a great advantage of SO(10) for approaching the problem of
understanding the quark and lepton masses and mixings is that SO(10)
relates all the types of fermions -- up quarks, down quarks, and
leptons -- and can provide a natural explanation of the smallness
of the KM angles.

\section{Conclusions}

The Dimopoulos-Wilczek mechanism has been shown to be a completely
natural way to achieve doublet-triplet splitting and a gauge
hierarchy. Not only can a Higgs sector be constructed for SO(10)
which implements this mechanism,$^{(1,15)}$ but it can be done so in such a way
that the superpotential is the most general consistent with some
symmetry. Moreover, even if Planck-scale physics induces
higher-dimension operators suppressed only by the dimensionally appropriate
powers of the Planck mass, it has been shown that local symmetry
can protect the DW mechanism from disruption by these operators.
It would appear that an invisible axion is an automatic consequence
of the DW mechanism if it is implemented in a fully natural way,
and that whatever local symmetries may be necessary to protect
the gauge hierarchy from Planck-scale effects also tend to protect
the axion. Finally, it has been shown how realistic and predictive
schemes for quark and lepton masses can be obtained within the
framework of the DW mechanism.

There are, of course, other attractive schemes of unification.
Each has strong and weak points. SUSY-SU(5) can be made natural
if 2/3 splitting is done using the ``missing-partner mechanism''.
The cost is the introduction of the somewhat high-rank Higgs
representations, ${\bf 50 + \overline{50} + 75}$. The main virtue
of SU(5) is that it is the smallest grand-unified group.
Flipped SU(5) (really $SU(5) \times U(1)$) has the great virtues
that the missing-partner mechanism can be implemented in a beautiful
and economical way, and that only small representations are
required. The main drawbacks are that the group is not simple,
so that the great accuracy of the $\sin  ^2 \theta _W$ prediction
has a less straightforward explanation, as has the relation
$m_b^0 = m_{\tau}^0$. Another intriguing possibility is the
group $SU(3) \times SU(3) \times SU(3)$,$^{(16)}$ which is suggested by
some superstring scenarios.

The good points of SO(10) unification are many and well-known. Some of them
have been mentioned in the introduction and in section 5 of this paper.
It would seem that a fully natural 2/3 splitting and gauge hierarchy in
SO(10) requires the Dimopoulos-Wilczek mechanism. We have shown that
this natural and realistic.

Finally, we have argued that ``weak suppression''
of higgsino-mediated proton decay can be achieved in much simpler
models than ``strong suppression'' (in the terminology of Ref. 1),
giving some reason to expect, in the context of SO(10), that
proton decay can be seen experimentally.
\newpage

\section*{References}

\begin{enumerate}
\item K.S. Babu and S.M. Barr, Phys. Rev. \underline{D48}, 5354 (1993).
\item S. Dimopoulos and F. Wilczek, Report No. NSF-ITP-82-07 (unpublished).
\item U. Amaldi, W. de Boer, and H. F\"{u}rstenau, Phys. Lett. \underline
{B260}, 447 (1991); P. Langacker and M.X. Luo, Phys. Rev. \underline
{D44}, 817 (1991); J. Ellis, S. Kelley, and D.V. Nanopoulos, Phys. Lett.
\underline{B260}, 131 (1991).
\item A. De Rujula, H. Georgi, and S.L. Glashow, Phys. Rev. Lett.
\underline{45}, 413 (1980); H. Georgi, S.L. Glashow, and M.Machacek,
Phys. Rev. \underline{D23}, 783 (1981); S.M. Barr, Phys. Lett.
\underline{112B}, 219 (1982).
\item See eg. ch IV of \underline{Unity of Forces in the Universe}, A. Zee
(1982, World Scientific Pub. Co. Inc., Singapore), and ``Grand Unified
Theories'', S.M. Barr, in \underline{Encyclopedia of Physics}, Second Edition,
ed. Rita G. Lerner and George L. Trigg (1991, VCH Pub. Inc.)
\item See eg., T. Banks, Physicalia \underline{12}, 19 (1990); J. Preskill,
Proceedings of the Int. Symp. of Gravity, the Woodlands, Texas (1992).
\item R. Peccei and H.R. Quinn, Phys. Rev. Lett. \underline{38}, 1440 (1977).
\item J.E. Kim and H.P. Nilles, Phys. Lett. \underline{B138}, 116 (1984).
\item S.M. Barr and D. Seckel, Phys. Rev. \underline{D46}, 539 (1992);
M. Kamionkowski and J. March-Russell, Phys. Rev. Lett. \underline{69},
1485 (1992);
R. Holman, S. Hsu, T. Kephart, E.W. Kolb, R. Watkins, and L. Widrow,
$ibid$. 1489 (1992);
K.S. Babu and S.M. Barr, Phys. Lett. \underline{B300}, 367 (1993).
\item S.M. Barr, Phys.Rev.Lett. \underline{64}, 353 (1990); Phys.Rev.
\underline{D42}, 3150 (1990).
\item M. Srednicki, Nucl.Phys. \underline{B202}, 327 (1982).
\item J. Preskill, M. Wise, and F. Wilczek, Phys. Lett. \underline{B120},
127 (1983); L. Abbott and P. Sikivie, ibid., 133; M. Dine
and W. Fischler, ibid., 137.
\item S.-Y. Pi, Phys. Rev. Lett. \underline{52}, 1725 (1984);
A. Linde, Phys. Lett. \underline{201B}, 437 (1988).
\item S.M. Barr, Phys. Rev. \underline{D24}, 1895 (1981) and
Ref. 10. For other recent papers proposing factorized quark and lepton
mass matrices see: B.S. Balakrishna, A.L. Kagan, and R.N. Mohapatra,
Phys. Lett. \underline{205B}, 345 (1988); K.S. Babu, B.S. Balakrishna, and
R.N. Mohapatra, $ibid$., \underline{B237}, 221 (1990); K.S. Babu and R.N.
Mohapatra, Phys. Rev. Lett. \underline{64}, 2747 (1990).
\item D. Lee and R.N. Mohapatra, Maryland Preprint UMD-PP-94-50 (1993).
\item G. Dvali and Q. Shafi, Bartol Preprint BA-93-47 (1993).
\end{enumerate}

\newpage
\noindent
{\large\bf Table I:} The particle content of the $SO(10) \times Z_3$ model
discussed in section 2.  Here $\omega \equiv e^{2 \pi i/3}$.
\vspace{.2in}

$$\begin{array}{c|ccccccccc} & F_I & T_1 & T_2 & A & A' &
A^{\prime\prime} & S & \bar{C} & C\\ \hline
Z_3 & \omega ^2 & \omega ^2 & 1 & \omega & \omega ^2 & 1 &
1 & \omega ^2 & \omega \\ & & & & & & & & & \\
SO(10) & 16 & 10 & 10 & 45 & 45 & 45 & 54 & \overline{126} & 126
\end{array}
$$
\vspace{2.0in}

\noindent
{\large\bf Table II:} The particle content of the $SO(10) \times U(1)$
model discussed in section 3.
\vspace{0.2in}

$$\begin{array}{c|cccccccccc} & T_1 & T_2 & T_3 & A & A' & A^{\prime\prime}
& S & (\bar{C} \cdot C) & P & \bar{P} \\ \hline
U(1) & -(\frac{3}{2}p + q) & \frac{3}{2}p & -\frac{1}{2}p & q &
-q & -p & 0 & p & p & -p \\ & & & & & & & & & & \\
SO(10) & 10 & 10 & 10 & 45 & 45 & 45 & 54 & (126 \cdot \overline{126}) &
1 & 1
\end{array}
$$

\newpage
\noindent
{\large \bf Figure Captions:}
\vspace{0.5in}

\noindent Fig.1: A graph
showing how in the model of section 2 proton decay is mediated
by color-triplet higgsinos. The amplitude is proportional to
$(M_2 + \lambda_2 \left< S \right>)/M_{GUT}$, which can be made
small. This is ``weak suppression'' in the terminology of Ref. 1.
\vspace{0.5in}

\noindent
Fig.2: A graph showing that if a term $MA^2$ exists in the superpotential
then symmetry will allow a $(T_1)^2$ term to exist as well.
\vspace{0.5in}

\noindent
Fig.3: A graph showing that if a $M(T_2)^2$ term exists in the
superpotential then symmetry will allow a $T_1 \cdot A \cdot A
\cdot T_1$ term. The related term $T_1 \cdot T_1$tr$(A^2)$ is
dangerous to the gauge hierarchy.
\vspace{0.5in}

\noindent
Fig.4: The analogue of Fig. 3 in a certain model with three {\bf 10}'s
of Higgs fields, showing that it is not necessary for terms containing
$(T_1)^2$ to arise at less than seventh order in superfields.
\vspace{0.5in}

\noindent
Fig.5: A graph that can induce a higher-dimension operator that
contributes to light quark and lepton masses. These operators
involve GUT-symmetry breaking and so can give realistic mass
relations.

\end{document}